\documentclass[times,preprint,showkeys,showpacs,longbibliography]{revtex4-1}
\usepackage{graphicx}
\usepackage{epstopdf}

\usepackage{amssymb}

\usepackage[figuresright]{rotating}

\begin{document}

\title{Centrality Fingerprints for Power Grid Network Growth Models}
\date{\today}

\author{Aleks Jacob Gurfinkel}
\email{agurfinkel@fsu.edu}

\author{Daniel A. Silva}
\author{Per Arne Rikvold}

\affiliation{Department of Physics, Florida State University, Tallahassee, Florida 32306-4350, USA}

\begin{abstract}
In our previous work, we have shown that many of the properties of the Florida  power grid are reproduced by deterministic network growth models based on the minimization of energy dissipation $E_\mathrm{diss}$. As there is no \em a priori \em best  $E_\mathrm{diss}$ minimizing growth model,  we here present a tool, called the ``centrality fingerprint," for probing the behavior of  different growth models. The centrality fingerprints are comparisons of the current flow into/out of  the network with the values of various centrality measures calculated at every step of the growth process. Finally, we  discuss applications to the Maryland power grid.
\end{abstract}


%89.75.Fb   \sep        % Self-organization:complex systems
%0.5.10.-a  \sep         % Computational techniques: statistical physics and nonlinear dynamics
%01.75.+m                 % Science and Society

\maketitle


\section{Background}
\label{sec:1}

\begin{figure}
\begin{center} \hspace*{-1.2cm}
\includegraphics[width=16cm]{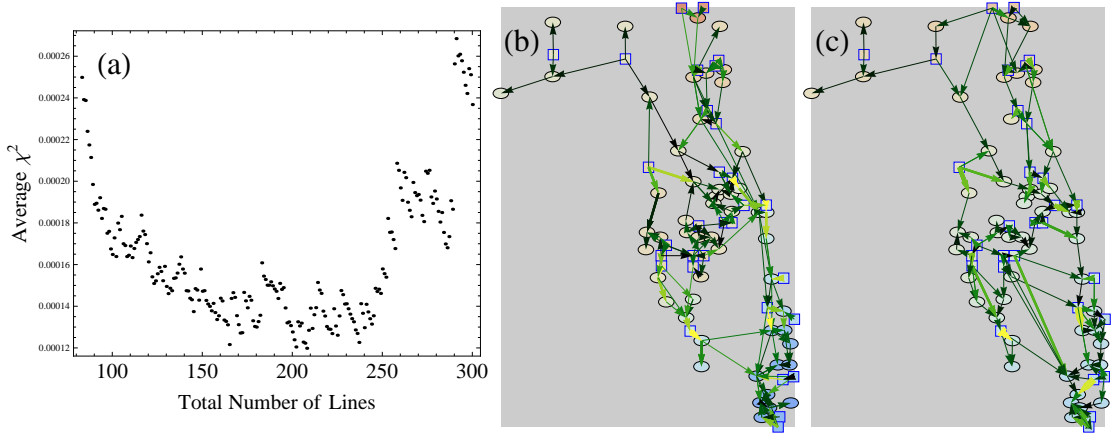}
\end{center}
\vspace{-2em}
\caption{ \label{fig:trip}Growth model based on real Florida vertex positions and in/out currents. (a) Illustration of the stopping criterion. The smallest difference ($\chi ^2$) between the growth-driving currents and exponential centrality is found when the number of lines $M=208$. See Fig. \ref{fig:finger}a for a generalized stopping criterion. (b) The FLG network. Arrowheads indicate direction of current flow along edges, while brighter arrow color indicates stronger currents. Thicker arrows indicate more parallel lines within multiedges. Red/blue vertex colors indicate high/low vertex potentials. (c) The grown network starting with the minimum spanning tree subgraph of the real Florida grid network and ending with $M=208$ lines according to the stopping criterion. }
\end{figure}

This work is concerned with the study of  power grids using the language and methods of complex network theory.  Since, in all their detail, today's electrical distribution grids are the largest engineered systems ever built\cite{backhaus2013getting}, our coarse-grained approach focuses exclusively on  high-voltage transmission lines, high-capacity generators, and switching and transmission substations. These three elements are represented by the edges and vertices of a complex network. Our primary test case is the real Florida power grid (FLG), whose $N=84$ vertex network is depicted in  Fig.~\ref{fig:trip}b.  Such  network models can be applied to test Intelligent Islanding strategies for limiting cascading blackouts\cite{Hamad20112}. 

Previous publications on this topic\cite{Hamad20112,Rikvold2012119} have focused on variants of a Monte Carlo ``cooling" model. In that approach, power lines are randomly connected between  loads and generators distributed randomly over a rectangular geography. This results in an unrealistically long total line length $L$, which is reduced by a Metropolis line-switching process, in which the Hamiltonian is equal to $L$ and the temperature is chosen to match the total length of lines in FLG. Thus, in the Monte Carlo power-grid models, both $L$ and $M$---the total number of lines---are explicitly matched. Further refinements in \cite{Xu2014130} have captured other features of FLG at the cost of explicitly matching the total edge resistance $R$, where the resistance of the edge $ij$ is  equal to $(\mbox{geographical distance between $i$ and $j$})/$ $(\mbox{number of lines between $i$ and $j$})$.

In an effort to illuminate the architecture of power grids, we have more recently sought to produce models that coincide with FLG in several key metrics, but without explicit matching. To this end, we have introduced a family of deterministic growth models that start with a minimal-length spanning tree over the vertices and add lines one by one according to a fixed rule.  The inspiration behind the rule comes from the behavior of general resistor networks with fixed current boundary conditions---currents flowing either in or out of the network at certain junctions. The problem is to solve for the internal currents flowing across the resistors of the network. In this situation, of all the current flows consistent with Kirchhoff's Current Law, the one realized by nature is that which minimizes the total energy dissipation $E_\mathrm{diss}\equiv \sum_{e\,\in\,\mathrm{edges}}I^2_eR_e$\cite{doyle}. One can apply this kind of dissipation optimization concept to any network that features in and out flows. In the present case---as in our previously published work---the AC power-distribution problem of the electrical grid is recast as a DC current-flow problem. The current flowing into the power grid network at a vertex is proportional to the generating capacity of a corresponding power plant. Analogously, the current flowing out of the network at a vertex is the corresponding load power consumption. Naturally, the energy dissipation of the grid will be highly sensitive to the distribution of these currents, and since this is the motivation behind our growth model, we call these in/out current values the \em growth-driving currents\em.

Choosing appropriate values for these currents is vital to the success of the model. Data for the generator capacities (the  positive growth-driving currents) are generally available from power-plant management. On the other hand, we must resort to estimation for the load power consumptions (negative growth-driving currents). In making these estimates, our guiding assumption is that topologically important load vertices will have relatively higher  out-currents. In the next section we sharpen the concept of topological ``importance," identifying it with network centralities found in the literature\cite{Newman}. Any free parameters introduced in our centralities must be calibrated to match the (known) generator data. The best agreement is found using the exponential centrality at $T\approx2.4$ (see Sec.~\ref{sec2}). We note that the growth-driving generator and load currents are normalized to $1$ and $-1$, respectively.

With the intuitively plausible $E_\mathrm{diss}$ optimization principle in hand, we may proceed to define various rules for choosing lines in the growth model. The best results (for the metrics in Table~\ref{tab:res}) have been found with the ``Cost/Benefit" model, in which the added line is chosen to minimize $\Delta L / |\Delta E_\mathrm{diss}|$. Here, the length of the line is associated with the cost, while the benefit is the \em drop \em in $E_\mathrm{diss}$. A rival model captures the same idea by focusing on the total (rather than marginal) cost and benefit, minimizing $E_\mathrm{diss}L^2$.  Additionally, we explore power-grid networks that are not cost-constrained:  we choose the line that minimizes $E_\mathrm{diss}$ at every step--- this we call the ``pure $E_\mathrm{diss}$ minimization" model.

The only element that remains to be specified is a stopping criterion for the network growth. The results reported here correspond to stopping when the $\chi^2$ between the growth-driving currents and the exponential centrality measure (with the parameter $T=2.4$) is minimal. This is described further in the the next section. The results of the growth model for real Florida generator/load geographies are reported in Fig.~\ref{fig:trip}. Here, the starting point of the growth is the minimal-length vertex-spanning tree subgraph of FLG. The results for certain key metrics are listed in Table~\ref{tab:res}. There, $C$ is the clustering coefficient\cite{Newman} and $e$ values are the network mixing patterns---the fraction of power lines with ends on different vertex types\cite{NewmanMixing}. We stress the lack of explicit matching for any of these metrics.

\begin{table*}[ht]

\setlength{\tabcolsep}{6pt}
\caption{\label{tab:res} Properties of the growth model for FLG. }
\begin{tabular}{lllllllllll}
\hline
Network        & $E_\mathrm{diss}$ & $M$ & \#  edges & \# multiedges & \multicolumn{1}{c}{$C$} & $L/N$ & $R$ & $e_{gg}$ & $e_{gl}$ & $e_{ll}$ \\\hline
FLG            & 0.029             & 200 & 137       & 48            & 0.21                    & 1.09                & 139. & 0.085    & 0.263    & 0.39     \\
Growth Model    & 0.013             & 208 & 130       & 43            & 0.22                    & 1.11                & 129. & 0.087    & 0.36     & 0.20     \\
  
\end{tabular}

\end{table*}

\vspace{-2em}

\section{Centrality fingerprints}\label{sec2}

\begin{figure}

\begin{center} \hspace*{-1.2cm}
\includegraphics[width=18cm]{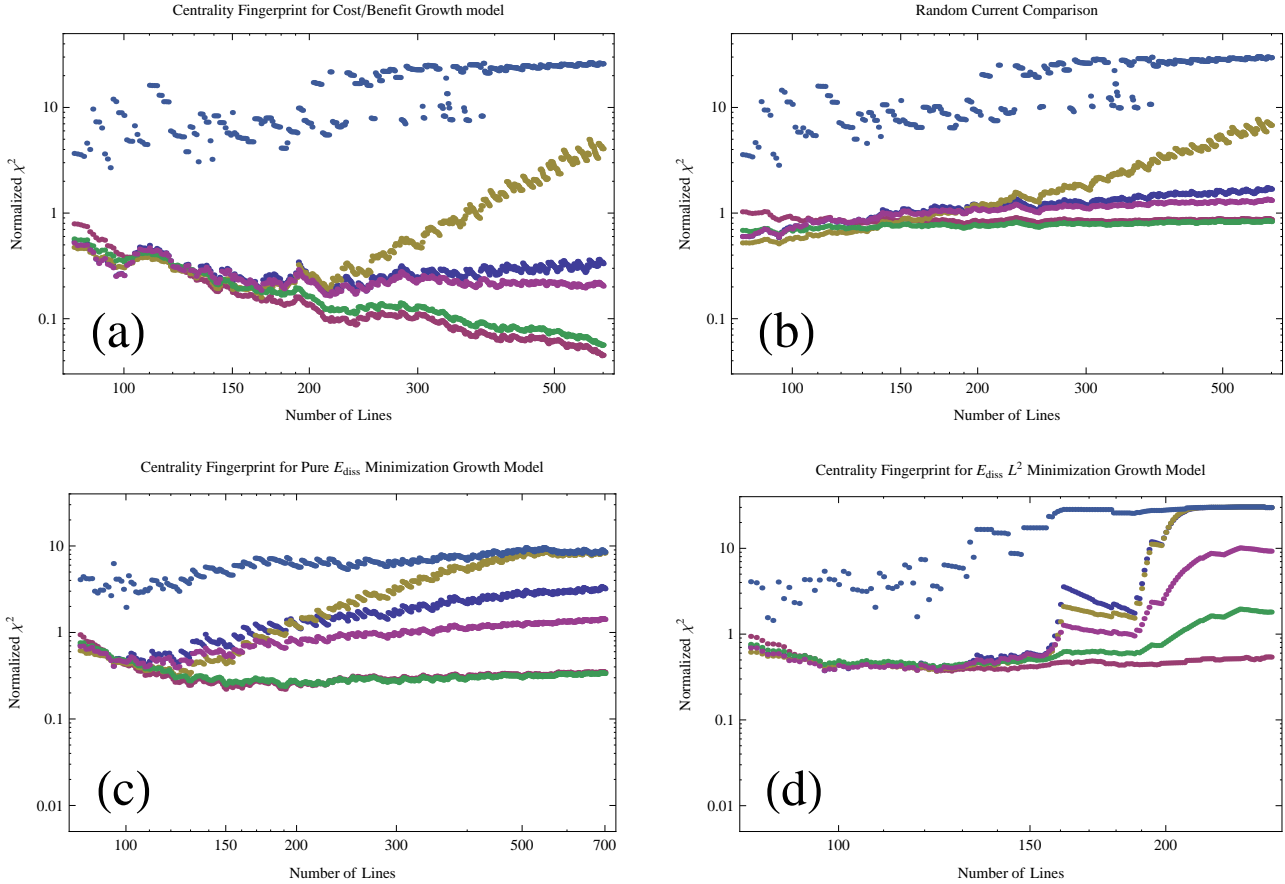}
\end{center}

\caption{ \label{fig:finger}
Centrality fingerprints for three $E_\mathrm{diss}$-based growth models. The centralities shown are eigenvector (light blue), exponential (gold), normalized exponential (dark blue), $A^2$ (purple), degree (green), and ``Ping-Pong" (red-purple). The preceding list is in descending order of $\chi^2$ for 500 lines in (a). Part (b) features a null-hypothesis result, where the growth-driving currents in the definition of $\chi^2$ are replaced by  randomly chosen currents. The centralities of primary interest  have $\chi^2$ values that stabilize near 1, the value for uncorrelated random currents.
 }
\end{figure}

We have remarked that the appropriate choice of centrality measure is critical to the success of the growth model. Many of the most common centralities can be written\cite{estrada2011structure} in the form $\left(\sum _{n=0}^{\infty } \mathbf{A}^nf_n\right)|1\rangle $, where  $|1\rangle $ is the column vector with all entries equal to 1, and $f_n$ is a function that assigns  weights to different powers of the network's adjacency matrix $\mathbf{A}$. To get insight into this formula, note that $(\mathbf{A}^n)_{ij}$ is equal to the number of paths of length $n$ from vertex $i$ to $j$. (Here the parallel lines that make up a multiedge contribute to distinct paths.) The $i$ component of the vector $\mathbf{A}^n\left|1\right>$ is then the total number of paths of length $n$ starting on vertex $i$. Our assumption is  that an ``important" vertex will be the endpoint of many paths. Furthermore, we wish to weight short paths more than long, winding paths; this can be accomplished by choosing a function $f_n$ that falls off appropriately fast.

Here we consider five centrality measures of this form. The simplest is \em degree centrality\em, proportional to the number of lines incident on each vertex and calculated from $\mathbf{A}|1\rangle$.  (We also consider $\mathbf{A}^2|1\rangle$.) The \em eigenvector centrality\em, proportional to the principal eigenvector of $\mathbf{A}$, is obtained from letting $f_n=0$ for all $n\neq n^\prime$ while $n^\prime\to \infty$. The \em exponential centrality\em, with adjustable parameter $T$, is given by $f_n = 1/(n! T^n )$. Note that, after many growth steps,  $\mathbf{A}$ will be large and higher-power terms will dominate, whence we recover eigenvector centrality. To address this issue, we also define the \em normalized exponential centrality \em with $f_n=(1/n! (T M)^n)$.

Recall that we have already employed  centrality measures to estimate the (negative) growth-driving currents. Here, we use them for a second purpose: to probe the extent to which the growth model seeks to reproduce the growth-driving currents. This can be quantified by a normalized $\chi^2$ measure equal to $N_\chi^{-1}\sum_{v\,\in\,\mathrm{vertices}}(\mathrm{current}(v) - \mathrm{centrality}(v))^2$. (We have chosen the normalizing factor $N_\chi^{-1}$ so that the maximum possible $\chi^2$ is approximately 100, while $\chi^2=1$ is obtained as the average value for uncorrelated random in/out currents satisfying our normalization requirements.) The $\chi^2$ can be calculated for every step in the growth to create a dynamical picture of the growth process. Of course, we do not expect  $\chi^2$ to have the same tendencies for every centrality.  We propose that comparing many different $\chi^2$s for a given growth model can give us (i) insight into the appropriateness of the centrality measure and (ii) a unique illustration of the growth model's behavior. We call such a comparison a ``centrality fingerprint," and display examples in Fig.~\ref{fig:finger}  for the three growth model variants previously discussed.

Comparing Figs.~\ref{fig:finger}a, \ref{fig:finger}c, and \ref{fig:finger}d, we  note the marked qualitative differences in the $\chi^2$ curves.  Though in each plot, the un-normalized exponential centrality is rising to meet the eigenvector centrality, the \em normalized \em exponential (along with the $\mathbf{A}^2$ centrality) stabilizes only  in the  Cost/Benefit model. Furthermore, only the Cost/Benefit model admits $\chi^2$s that continually fall as the network is grown. This suggests that the Cost/Benefit model asymptotically matches these centralities to the growth-driving currents, and thus that those centralities are closely related to that growth model. 

The lowest $\chi^2$ obtained for the Cost/Benefit model has not yet been introduced. This is the ``Ping-Pong" centrality, which we developed to capture the relationship between sources and sinks. This centrality marks generators as important if they have many paths to loads and vice versa. This is accomplished by simulating  many random walks on the network, alternating between  absorbing on generators and loads. The Ping-Pong centrality can be expressed as the principal eigenvalue of a certain walk matrix, a form which allows us to prove that it reduces to the degree centrality in bipartite networks; this is echoed in the similarity of the $\chi^2$ for degree and Ping-Pong centralities  in Figs.~\ref{fig:finger}a and  \ref{fig:finger}c. 

 Finally, we  note that in Figs.~\ref{fig:finger}a and \ref{fig:finger}c we see minima in certain $\chi^2$ values at a number of lines ($\approx 200$) corresponding to the number of lines in the real Florida grid. Any of these minima could be plausibly  identified as stopping points for the network growth model.

In the preceding considerations, we have compared growth models using the real Florida geography and generator capacities. It is reasonable to expect that varying these conditions significantly may lead to different behavior. For example, the power grid of the state of Maryland\cite{maryland} has different network characteristics than  FLG, including a significantly lower clustering coefficient. Our Cost/Benefit model run with the Maryland data creates a centrality fingerprint very different from that in Fig.~\ref{fig:finger}a. A possible explanation is that none of the centralities considered here provide a good estimate for the Maryland generating capacities, leading to an invalid extrapolation to the load power demands. We leave it for future research to find appropriate centrality measures for the Maryland and other grids.

\begin{acknowledgements}
Supported in part by NSF Grant Nos. DMR-1104829 and HRD-1143070.
\end{acknowledgements}

\end{document}